\documentclass[11pt]{article}
\usepackage{amsfonts}
\usepackage{amsthm}

\usepackage{amsmath,amssymb}

\def\be{\begin{equation}}
\def\ee{\end{equation}}
\def\bea{\begin{eqnarray}}
\def\eea{\end{eqnarray}}

\newcommand{\sect}[1]{\setcounter{equation}{0}\section{#1}}

\newcommand{\bq}{\mathbf{q}}
\newcommand{\bp}{\mathbf{p}}

\newcommand{\btq}{ \mathbf{q}}

\newcommand{\tq}{ {q}}

 \newcommand{\kk}{\kappa}
   
  \newcommand{\Om}{\Omega}

  \newcommand{\nonn}{ {\cal H}}

\newcommand{\iii}{ {\cal I}}

\newcommand{\ele}{ {\cal L}}

\newcommand{\KK}{ {\cal K}}

   \def\M#1{{ (#1)}}

 \def\k{{\kappa}}

\def\1{\'{\i}}                           

  \def\>#1{{\mathbf#1}}

   \def\m{\mu}

\begin{document}

\thispagestyle{empty}

\

 \vskip2cm

\begin{center}

 {\Large{\bf {An integrable H\'enon--Heiles system\\[6pt]}}}

 {\Large{\bf {on the sphere and the hyperbolic plane}}}

\medskip 
\medskip 
\medskip

{\sc \'Angel Ballesteros$^\dagger$, Alfonso Blasco$^\dagger$, Francisco J.~Herranz$^\dagger$ and Fabio Musso$^{\ddagger}$}

\medskip

{$^\dagger$ Departamento de F\'\i sica,  Universidad de Burgos,
E-09001 Burgos, Spain  \\
$^\ddagger$ Istituto Comprensivo ``Largo Dino Buzzati" I-00144, Rome, Italy \\
}
\medskip 

\noindent
 E-mail: {\tt   angelb@ubu.es,  ablasco@ubu.es,  fjherranz@ubu.es, fmusso@ubu.es}

\end{center}

  \medskip 
\bigskip
\bigskip

\begin{abstract} 
  We  construct a constant curvature analogue  on the two-dimensional sphere ${\mathbf S}^2$  and  the hyperbolic  space ${\mathbf H}^2$ of the integrable H\'enon--Heiles   Hamiltonian $\mathcal{H}$ 
   given by
$$
\mathcal{H}=\dfrac{1}{2}(p_{1}^{2}+p_{2}^{2})+ \Omega \left(  q_{1}^{2}+ 4 q_{2}^{2}\right) +\alpha \left(
q_{1}^{2}q_{2}+2 q_{2}^{3}\right) ,
$$
where  $\Omega$ and  $\alpha$   are real constants. The curved integrable Hamiltonian $\mathcal{H}_\kappa$ so obtained depends on a parameter $\k$ which is just the   curvature of the  underlying space, and is such that the Euclidean H\'enon--Heiles system $\mathcal{H}$ is smoothly obtained in the zero-curvature limit $\k\to 0$. On the other hand, the Hamiltonian $\mathcal{H}_\kappa$ that we propose can be regarded as an integrable perturbation of a known curved integrable $1:2$ anisotropic oscillator. We stress that in order to obtain the curved  H\'enon--Heiles  Hamiltonian $\mathcal{H}_\kappa$,  the preservation of the full integrability structure of the flat Hamiltonian $\mathcal{H}$ under the deformation generated by the curvature will be imposed. In particular, the existence of a curved analogue of the full Ramani--Dorizzi--Grammaticos  (RDG)  series $\mathcal{V}_{n}$ of integrable polynomial potentials, in which the flat H\'enon--Heiles   potential can be embedded, will be essential in our construction. Such infinite family of curved RDG potentials $\mathcal{V}_{\k, n} $ on ${\mathbf S}^2$  and ${\mathbf H}^2$  will be also explicitly presented.
 \end{abstract}

\bigskip\bigskip\bigskip\bigskip


\bigskip


\newpage


\sect{Introduction}

The problem considered in this paper can be stated in generic terms as follows: given a certain   Liouville integrable Hamiltonian system on the two-dimensional (2D)  Euclidean space
\be
\mathcal{H}=\mathcal{T} + \mathcal{V}=\dfrac{1}{2}(p_{1}^{2}+p_{2}^{2}) + \mathcal{V}(q_1,q_2),
\nonumber
\ee
and whose integral of the motion is given by $\mathcal{I}(p_1,p_2,q_1,q_2)$,
find a one-parameter integrable generalization $\mathcal{H}_\kappa$ of this system of the form
\be
\mathcal{H}_\kappa=\mathcal{T}_\kappa(p_1,p_2,q_1,q_2) + \mathcal{V}_\kappa(q_1,q_2),
\label{ansatz}
\ee
with integral of the motion given by $\mathcal{I}_\kappa(p_1,p_2,q_1,q_2)$ and fulfilling the following conditions:
\begin{itemize}

\item $\mathcal{T}_\kappa$ has to be the kinetic energy of a particle on a 2D space with constant curvature $\kappa$, namely the sphere \textbf{S}$^{2}$ ($\kappa>0$), the hyperbolic space \textbf{H}$^{2}$ ($\kappa<0$) and the Euclidean space \textbf{E}$^{2}$ ($\kappa=0$). To this aim we will use the curvature-dependent formalism introduced in~\cite{CK2,RS,lett,Voz}, and the explicit form of $\mathcal{T}_\kappa$ will be given afterwards  (it will obviously depend upon the particular
set of coordinates chosen).

\item The Euclidean Hamiltonian $\mathcal{H}$ has to be smoothly recovered in the zero-curvature limit $\kappa\to 0$, namely
$$
\mathcal{H}=\lim_{\kappa\to 0}{\mathcal{H}_\kappa},
\qquad
\mathcal{I}=\lim_{\kappa\to 0}{\mathcal{I}_\kappa}.
$$
In particular, since by construction $\mathcal{T}=\lim_{\kappa\to 0}{\mathcal{T}_\kappa}$, we have to impose that $\mathcal{V}=\lim_{\kappa\to 0}{\mathcal{V}_\kappa}$.

\end{itemize}

If these conditions are fulfilled, we shall say that $\mathcal{H}_\kappa$ is a {\em curved $\mathcal{H}$ system on the sphere and the hyperbolic space}. Note that, in principle, the uniqueness of this construction is not guaranteed, since different $\mathcal{V}_\kappa$ potentials (and their associated $\mathcal{I}_\kappa$ integrals) having the same $\kappa\to 0$ limit could be found. An example of such non-uniqueness has been explicitly given in~\cite{Non, Non2}, where the construction of  integrable curved analogues of the anisotropic oscillator potential was studied. Moreover, if the Hamiltonian $\mathcal{H}$  is superintegrable ({\em i.e.}~if another globally defined and functionally independent integral of the motion ${\KK}(p_1,p_2,q_1,q_2)$ does exist) then we could further impose the existence of the curved (and functionally independent) analogue ${\KK}_\kappa$ of the second integral. If we succeed in finding such second integral, we would obtain a {\em superintegrable curved} generalization of $\mathcal{H}$.

 In this respect,  it is worth recalling that the search of integrable potentials with quadratic integrals of motion on \textbf{S}$^{2}$ (and therefore admitting separation of variables) was formerly considered in~\cite{Wojc2,Bogoyavlensky}, and that a complete classification of superintegrable potentials (again with quadratic integrals) on 
\textbf{S}$^{2}$ and \textbf{H}$^{2}$ was further presented in~\cite{RS,Kalnins1,Kalnins2}.
More recently,  trajectory isomorphisms between integrable systems on the $N$D sphere and the $N$D Euclidean space have been established as a result of central projection in~\cite{Borisov2}.

We would like to stress that within  the framework here presented the   Gaussian curvature $\kappa$  of the space  enters as a  deformation parameter, and the curved systems $\mathcal{H}_\kappa$ can be thought of as integrable perturbations of the flat ones $\mathcal{H}$ in terms of the curvature parameter $\kappa$. In this way, integrable Hamiltonian systems on \textbf{S}$^{2}$ ($\kappa >0$),  \textbf{H}$^{2}$ ($\kappa <0$) and \textbf{E}$^{2}$ ($\kappa =0$) can be simultaneously constructed and analysed.
This approach has been followed so far in order to  construct, mainly, analogues of the oscillator and Kepler--Coulomb systems on spaces with constant curvature (see~\cite{RS, lett, Voz, Non, Non2, CRMVulpi,  CRS, kiev, HBlettercoulomb, DiacuJDE,  Diacu1, Diacu2, DiacuM, Rastelli} and references therein). 
Other classical and modern results in this field, albeit without following such curvature-dependent approach, can be found in~\cite{Killing, Appell, Higgs, Leemon, Kozlov, Schrodingerdualb, Borisov1, Albouy,Bizyaev}.

 In particular, this paper is devoted to the construction of the first, to the best of our knowledge, example of a curved integrable H\'enon--Heiles system in the above mentioned sense.
We recall that the original H\'enon--Heiles Hamiltonian~\cite{HH} 
\be
{H}=\dfrac{1}{2}(p_{1}^{2}+p_{2}^{2}) + \dfrac{1}{2}(q_{1}^{2}+q_{2}^{2})+\lambda\left(
q_{1}^{2}q_{2}-\frac{1}{3}\,q_{2}^{3}\right)
\label{HHaut}\nonumber
\ee
is not Liouville-integrable, and provides an outstanding example of non-linear dynamical system that exhibits chaotic behaviour (see, for instance, \cite{Tabor, Gutzwiller, BoPu}). Nevertheless, its multiparametric generalization
\be
 {H}=\dfrac{1}{2}(p_{1}^{2}+p_{2}^{2})+ \Omega_{1}  q_{1}^{2}+\Omega_{2} q_{2}^{2}+\alpha \left(
q_{1}^{2}q_{2}+\beta q_{2}^{3}\right) ,
\label{hhmulti}
\ee
was soon proven to be Liouville integrable for three specific sets of values of the real parameters $\Omega_{1},\Omega_{2}$ and $\beta$ (see~\cite{BSV, CTW, GDP, HietarintaRapid, Fordy83, Wojc, SL, FordyHH, Sarlet, RGG, Hindues, Pickering}  and references therein), thus giving rise to the following distinguished families of integrable H\'enon--Heiles Hamiltonians:

\begin{itemize}

\item The Sawada--Kotera system: $\beta=1/3$ and $\Omega_{2}=\Omega_{1}$, which is separable in rotated Cartesian coordinates.

\item  The Korteweg--de Vries (KdV) system: $\beta=2$  with $\Omega_{1}$ and $\Omega_{2}$ arbitrary, which is separable in parabolic coordinates.

\item The Kaup--Kuperschdmit system: $\beta=16/3$ and $\Omega_{2}=16\,\Omega_{1}$,  whose integral of the motion is quartic in the momenta.

\end{itemize}

Notice that  the approach that we will follow is based on the fact that the H\'enon--Heiles Hamiltonian~\eqref{hhmulti} can be interpreted as a cubic perturbation of the anisotropic oscillator potential with frequencies $\omega_1^2=2\Omega_1$ and $\omega_2^2=2\Omega_2$.
Furthermore,  if we consider a particular KdV   system (\ref{hhmulti}) by setting  $\Omega_2=4\Omega_1$  (so with $\beta=2$), the above three integrable H\'enon--Heiles Hamiltonians can be regarded, respectively, as integrable cubic perturbations of the $1:1$, $1:2$ and $1:4$ superintegrable  anisotropic  oscillators~\cite{Jauch, Tempesta}, namely
\bea
&& {H}^{\rm SK}=\dfrac{1}{2}\left(p_{1}^{2}+p_{2}^{2} \right)+ \frac {1}{2}\,\omega^2 \left( q_{1}^{2}+   q_{2}^{2}\right) +\alpha \left(
q_{1}^{2}q_{2}+\frac 13\, q_{2}^{3}\right) , \label{wqa}\\
&& {H}^{\rm KdV}=\dfrac{1}{2}(p_{1}^{2}+p_{2}^{2})+ \frac {1}{2}\,\omega^2 \left( q_{1}^{2}+  4 q_{2}^{2}\right) +\alpha \left(
q_{1}^{2}q_{2}+2 q_{2}^{3}\right) ,\label{aa}\\
&& {H}^{\rm KK}=\dfrac{1}{2}(p_{1}^{2}+p_{2}^{2})+ \frac {1}{2}\,\omega^2 \left( q_{1}^{2}+   16 q_{2}^{2}\right) +\alpha \left(
q_{1}^{2}q_{2}+\frac {16} 3\, q_{2}^{3}\right) ,\label{wqc}
\eea
 such that $\omega^2=2\Omega_1$.

In this paper we will present the generalization on the 2D sphere \textbf{S}$^{2}$ and the   hyperbolic (or Lobachevski) space  \textbf{H}$^{2}$ of the integrable H\'enon--Heiles Hamiltonian of KdV type (\ref{aa}). We stress that the system ${H}^{\rm KdV}$  is endowed with additional and outstanding integrability properties, that will be essential in order to construct its curved analogue. In particular, the Euclidean Hamiltonian ${H}^{\rm KdV}$  (\ref{aa}) is deeply related with the so-called  {\em  Ramani--Dorizzi--Grammaticos  (RDG) series}~\cite{RDGprl, Hietarinta} of integrable homogeneous polynomial potentials of degree $n$ on the plane that are separable in parabolic coordinates~\cite{Wojc,FF}. They are given by
\be
\mathcal{V}_{n}(q_1,q_2)=\sum\limits_{i=0}^{[\frac{n}{2}]}2^{n-2i}\dbinom{n-i}{i}q_{1}^{2i}q_{2}^{n-2i}
\, ,\qquad n=1,2,\dots
\label{Ram1}
\ee
and, due to their separability, all these potentials  can be freely superposed by preserving integrability.
The first RDG potentials are explicitly given by
\bea
&& \mathcal{V}_{1}=2q_{2} ,\nonumber \\ 
&& \mathcal{V}_{2}=q_{1}^2+ 4q_{2}^2 ,\nonumber \\
&& \mathcal{V}_{3}=4q_{1}^2q_{2}+ 8q_{2}^3 ,\nonumber \\
&& \mathcal{V}_{4}=q_{1}^4+ 12 q_{1}^2 q_{2}^2 + 16 q_{2}^4  .
\nonumber
\eea
Therefore, by considering the system
\be
\mathcal{H}=\dfrac{1}{2}(p_{1}^{2}+p_{2}^{2})+\alpha_{2}\mathcal{V}_{2}+\alpha_{3}\mathcal{V}_{3}
\nonumber
\ee
with
$
\alpha_{2}=\omega^2/2
$
and
$
\alpha_{3}={\alpha}/{4}
$, we get  the KdV H\'enon--Heiles Hamiltonian (\ref{aa}). As a byproduct,  all the elements of the RDG series~\eqref{Ram1} can be considered as integrable polynomial perturbations of arbitrary degree $n$ of such H\'enon--Heiles system of KdV type. 
Moreover, the ability of the system ${H}^{\rm KdV}$ in order to admit integrable perturbations is enhanced by the fact that certain {\em rational} integrable perturbations can also be  superposed to the RDG series by preserving the integrability of the perturbed Hamiltonian (see~\cite{FF,HonePLA,tesis,Annals10} and references therein). All these strong integrability properties are indeed connected with the fact that ${H}^{\rm KdV}$ is related with a certain reduction of the KdV equation~\cite{FordyHH, HonePLA, Tondo, Conte, HoneIP}.

It is important to stress that the strategy that we will follow in order to construct the curved H\'enon--Heiles system is based on the preservation of the full integrability structure that we have already described, {\em i.e.}, we will consider the curved  H\'enon--Heiles system as an integrable perturbation of the $1:2$ superintegrable curved oscillator already studied in~\cite{RS, Non, Non2}, and  we will assume that such curved ${H}^{\rm KdV}$ system should be also embedded within the appropriate curved RDG series of potentials.

The paper is structured as follows.  The next Section is devoted to recall all the integrability features of the (Euclidean) integrable H\'enon--Heiles  system (\ref{aa}) and its related  RDG series (\ref{Ram1}). In Section 3, we briefly  review the two  coordinate systems on \textbf{S}$^{2}$ and  \textbf{H}$^{2}$ that will be useful in  order   to work out  the   curved version of these systems, namely, the ambient (or Weierstrass) and Beltrami (projective) canonical variables. Section 4 presents the main results of the paper. Firstly, by starting from the known curved $1:2$ superintegrable oscillator~\cite{RS, Non}, the  explicit expression for the full curved  integrable RDG series of potentials $\mathcal{V}_{\k, n}$ on \textbf{S}$^{2}$ and  \textbf{H}$^{2}$ are obtained. Secondly,
the corresponding  integrable curved H\'enon--Heiles Hamiltonian will be defined as a suitable superposition of the $n=2$ and $n=3$ terms in such series. The final Section deals with some comments and open problems.

\sect{The flat integrable H\'enon--Heiles Hamiltonian}
By setting   $\beta=2$ in (\ref{hhmulti}) we recover the KdV H\'enon--Heiles Hamiltonian in terms of Cartesian canonical variables on \textbf{E}$^{2}$  with $\{q_{i},p_{j}\}=\delta_{ij}$, namely
\begin{equation}
 {H} =\dfrac{1}{2}(p_{1}^{2}+p_{2}^{2})+\Omega_{1}q_{1}^{2}+\Omega_{2}q_{2}^{2}+\alpha\left(q_{1}^{2}q_{2}+2 q_{2}^{3}\right) .\label{KdVflata}
\end{equation}
We recall that in~\cite{Wojc}  this Hamiltonian was shown to be separable in terms of shifted parabolic coordinates and, furthermore,  the solution of the corresponding Hamilton--Jacobi equation was also obtained.
The corresponding constant of motion  is quadratic in the momenta, and reads
\be
 {I} = (4\Omega_{1}-\Omega_{2})\left(
\dfrac{p_{1}^{2}}{2}+\Omega_{1}q_{1}^{2}
\right)+\alpha\left(
p_{1}(q_{1}p_{2}-q_{2}p_{1})+q_{1}^{2}\left[2\Omega_{1}q_{2}+
\dfrac{\alpha}{4}(q_{1}^{2}+4 q_{2}^{2})
\right]
\right) .
\nonumber
 \ee

As mentioned above,  the   Hamiltonian  (\ref{KdVflata}) can be regarded as an {\em integrable cubic perturbation} of the $1:2$ anisotropic oscillator when    $\Omega_2=4\Omega_1\equiv 4\Omega$, so leading to  (\ref{aa}) with $ \omega^2=2\Omega $. Indeed, such anisotropic oscillator will be our starting point in order to construct the corresponding curved generalization on \textbf{S}$^{2}$ and \textbf{H}$^{2}$.  Hence the explicit Hamiltonian and the 
integral of motion that we will consider will be
\bea
&&\mathcal{H} =\dfrac{1}{2}(p_{1}^{2}+p_{2}^{2})+\Omega \left( q_{1}^{2}+4 q_{2}^{2}\right) +\alpha\left(q_{1}^{2}q_{2}+2 q_{2}^{3}\right) ,\label{KdVflat}\\[2pt]
&&   \mathcal{I} =  
p_{1}(q_{1}p_{2}-q_{2}p_{1})+q_{1}^{2}\left[2\Omega q_{2}+
\dfrac{\alpha}{4}\left(q_{1}^{2}+4 q_{2}^{2}\right)
\right] .
 \label{I2KdVflat}
 \eea

Now let us recall that the RDG potentials $\mathcal{V}_{n}(q_1,q_2)$~\cite{RDGprl, Hietarinta}
 are integrable homogeneous polynomial potentials of degree $n$   given by~\eqref{Ram1}. It is well  known that,  for a given $n$, the integral of the motion $\mathcal{L}_n$ for the RDG potential $\mathcal{V}_{n}$ contains the $\mathcal{V}_{n-1}$ potential. Explicitly, we have that
\be
 \{\mathcal{H}_{n},\mathcal{L}_n\}=0,
 \nonumber
\ee
 where
 \be
\mathcal{H}_{n}=\dfrac{1}{2}(p_{1}^{2}+p_{2}^{2})+\alpha_{n}\mathcal{V}_{n}, \quad \ 
\mathcal{L}_n=p_{1}(q_{1}p_{2}-q_{2}p_{1})+ \alpha_{n} q_{1}^{2}\mathcal{V}_{n-1} ,\quad\  
\label{Rflat}
\ee
and the recurrence works provided that we have defined $
 \mathcal{V}_{0}:=1$. Moreover, it is worth stressing that 
 all the RDG potentials can be freely superposed without losing   integrability \cite{Hietarinta,tesis,Annals10},  that is, the Hamiltonian 
\be
\mathcal{H}_{\M M}=\dfrac{1}{2}\left(
p_{1}^{2}+p_{2}^{2}
\right)+\sum\limits_{n=1}^M
\alpha_n \mathcal{V}_{n} =\dfrac{1}{2}\left(p_{1}^{2}+p_{2}^{2}\right)+\sum\limits_{n=1}^{M}\sum\limits_{i=0}^{[\frac{n}{2}]}\alpha_{n}2^{n-2i}\dbinom{n-i}{i}q_{1}^{2i}q_{2}^{n-2i} ,
\label{bk}
\ee
is endowed with a (quadratic in the momenta)  integral of the motion   given by
\bea
\mathcal{L}_{\M M}\!\!&=&\!\! p_{1}(q_{1}p_{2}-q_{2}p_{1})
+q_{1}^{2}\sum\limits_{n=1}^{M}\alpha_{n}\mathcal{V}_{n-1} \cr
\!\!&=&\!\! p_{1}(q_{1}p_{2}-q_{2}p_{1})
+q_{1}^{2}\left(
\sum\limits_{n=1}^{M}\sum\limits_{i=0}^{[\frac{n-1}{2}]}\alpha_{n}2^{n-1-2i}\dbinom{n-1-i}{i}q_{1}^{2i}q_{2}^{n-1-2i}\right).\label{I2Ramani}
\eea

As a consequence, the relationship among the   Hamiltonian (\ref{KdVflat}) and its integral of motion  (\ref{I2KdVflat})  with the RDG potentials (\ref{Ram1}) is immediately established through the superposition (\ref{bk}) and (\ref{I2Ramani})  by setting
\be
\alpha_1=0,\qquad 
\alpha_2=\Omega,\qquad  \alpha_3= \alpha/4 , 
\label{xz}
\ee
which yields  
\bea
&& \mathcal{H} \equiv \mathcal{H}_{\M 3} =\dfrac{1}{2}(p_{1}^{2}+p_{2}^{2})+\alpha_{2} \mathcal{V}_{2}+\alpha_{3} \mathcal{V}_{3} ,
\nonumber\\
&&  \mathcal{I} \equiv \mathcal{L}_{\M 3}=  
p_{1}(q_{1}p_{2}-q_{2}p_{1})+q_{1}^{2}\left(\alpha_2\mathcal{V}_{1} +
 {\alpha_3} \mathcal{V}_{2}   
\right). 
\nonumber
\eea
Therefore,  the potentials $\mathcal{V}_{2}$  and $\mathcal{V}_{3}$ define the H\'enon--Heiles system, meanwhile the integral  $\mathcal{I}$   contains the linear   $ \mathcal{V}_{1}$ as well as the  quadratic $ \mathcal{V}_{2}$ RDG potentials. As we will see in the sequel, this pattern (together with an appropriate choice of the coordinates on ${\mathbf S}^2$ and ${\mathbf H}^2$) will be essential for the construction of   the corresponding curved counterpart  of this  system.


\sect{Geometry and geodesic dynamics on ${\mathbf S}^2$ and ${\mathbf H}^2$}

A useful description of the dynamics of integrable systems   on constant curvature spaces is provided by a Lie-algebraic approach based on the following one-parameter family of 3D real Lie algebras   $\mathfrak{so}_{\kappa}(3)$ (see~\cite{Non,Non2} for details):
\be
  [J_{12},J_{01}]=J_{02},\qquad [J_{12},J_{02}]=-J_{01},\qquad [J_{01},J_{02}]=\kk J_{12}  , \label{ca}
 \ee
 where $\k$ is a real parameter.   The Casimir invariant of this algebra, which comes from the  Killing--Cartan form, reads
\be
  {\cal C}=J_{01}^2+J_{02}^2+\kk J_{12}^2.
 \label{caa}
 \ee
 Now, the   2D  Riemannian spaces of constant curvature  are defined as the homogeneous spaces $  {\rm  SO}_{\k}(3)/{\rm  SO}(2) $ where
  ${\rm SO}_{\k}(3)$ is the Lie group of   $\mathfrak{so}_{\kappa}(3)$ and ${\rm  SO}(2) $  is the isotopy subgroup generated by $J_{12}$. Therefore 
  $J_{12}$ corresponds to 
the generator of rotations leaving the origin $O$ invariant,     meanwhile $J_{01}$ and $J_{02}$ are the generators of translations moving $O$ along two basic directions. In this approach the parameter $\k$ is just the Gaussian {\em curvature} of the space and according to its value we  find:
 $$
\begin{array}{lll}
\kk>0:\ \mbox{Sphere}&\qquad \kk=0:\ \mbox{Euclidean plane}&\qquad \kk<0:\ \mbox{Hyperbolic space}\\[2pt]
  {\mathbf S}^2={\rm SO}(3)/{\rm  SO}(2)&\qquad  {\mathbf E}^2={\rm  ISO}(2)/{\rm  SO}(2)&\qquad {\mathbf H}^2= {\rm  SO}(2,1)/{\rm SO}(2)
\end{array}
$$
 
 The spaces ${\mathbf S}^2$ and ${\mathbf H}^2$ can be  embedded in a 3D  linear space $\mathbb R^3=(x_0,x_1,x_2)$ where the {\em ambient} (or Weierstrass) coordinates must fulfil the constraint 
 \begin{equation}
\Sigma_\k: x_{0}^{2}+\kappa (x_{1}^{2}+x_{2}^{2})=1,\label{Wcoor}
\end{equation}
  such that the origin is $O=(1,0,0)\in \mathbb R^3$. 
Now, we apply a central projection with pole  
$(0,0,0)\in \mathbb R^{3}$ from the ambient coordinates $(x_0,x_1,x_2)\in \mathbb R^3$ to the 2D projective space expressed in terms of 
   {\em Beltrami}   coordinates $ (\tq_1,\tq_2)\in \mathbb R^2$  defined by
$$
(x_0, x_1,x_2)\in\Sigma_\k \longrightarrow (0,0,0)+\m\,
(1,\tq_1,\tq_2) \in\Sigma_\k.
$$
By taking into account the constraint (\ref{Wcoor}), the projection yields the relationships between the ambient and the projective coordinates
\be
x_0=\m=\frac{1}{\sqrt{1+ \kk \btq^2}},\qquad
\>x=\m\, \btq=\frac{\btq}{\sqrt{1+ \k \btq^2}},\qquad \>q=\frac{\>x}{x_0},
\label{Beltr}
\ee
where from now on    we assume the   notation   $\>x=(x_1,x_2)$,   $\bq=(q_1,q_2)$ and  $\bp=(p_1,p_2)$ together with
$$
\>x^2=x_1^2+x_2^2,\qquad \bq^2=q_1^2+q_2^2,\qquad \bp^2=p_1^2+p_2^2,\qquad \bq\cdot \bp=q_1 p_1+ q_2 p_2  .
\label{bba}
$$

In terms of the Beltrami  canonical variables $(\bq,\bp)$,
a symplectic realization of  the Lie--Poisson analogue of the algebra $\mathfrak{so}_\kk(3)$ (\ref{ca})   is found to be~\cite{Non,Non2} 
 \be
J_{0i}= p_i+\kk (\bq\cdot\bp) q_i,   \quad i=1,2; \qquad J_{12}=q_1 p_2 - q_2 p_1 .
  \label{cd}
 \ee
Therefore, the free Hamiltonian $ {\cal T}_\k$,     providing  the kinetic energy term on these spaces,   is  directly  obtained from the Casimir (\ref{caa}) under the above realization, namely
 \be
 {\cal T}_\k\equiv\frac 12 {\cal C}=\frac 12 (J_{01}^2+J_{02}^2+\kk J_{12}^2)=\frac 12 \left(1+\kk \bq^2\right) \left(\bp^2+\kk(\bq\cdot\bp)^2 \right) ,
 \label{ce}
 \ee
 and this will be the (projective) kinectic energy term that we will consider for the Ansatz (\ref{ansatz}).
We stress that the flat/Euclidean limit $\k\to 0$ is always well defined in all the above expressions, thus leading to the Cartesian canonical variables on 
$ {\mathbf E}^2$ used in the previous Section. In particular, when $\k=0$ we have
$$
x_0=1,\qquad \>x = \btq,\qquad J_{0i}=p_i,\qquad J_{12}= q_1 p_2 - q_2 p_1,\qquad  {\cal T}= \frac 12 \bp^2.
$$

We recall that a complete review of   these  canonical variables   as well as other different ones (Poincar\'e and geodesic polar variables) together with  the relationships  among them have been presented  in \cite{Non2},  so as to provide a complete view of several possible descriptions of the dynamics of   Hamiltonian systems on ${\mathbf S}^2$ and ${\mathbf H}^2$.  Further results on geodesic motion   on spaces of constant curvature can be found in~\cite{RS, Voz,CRMVulpi, kiev}   (see also references therein).


\sect{The curved H\'enon--Heiles Hamiltonian}

Our construction of  a curved version for the   Hamiltonian  (\ref{KdVflat}) and its integral (\ref{I2KdVflat}) will have as a guiding principle the preservation of the full integrability structure of the Euclidean H\'enon--Heiles Hamiltonian. This implies  that we have to  obtain  the curved counterpart $ \mathcal{V}_{\k,n}$   of   the flat  RDG potentials  $ \mathcal{V}_{n}$ (\ref{Ram1}). The only known initial data for this construction will be the $ \mathcal{V}_{\k,2}$ potential, which will be given by the
generalization to  ${\mathbf S}^2$ and  ${\mathbf H}^2$ of the  Euclidean  $1:2$  anisotropic oscillator potential $ \mathcal{V}_{2}=q_{1}^2+ 4q_{2}^2$, which was firstly introduced in~\cite{RS} and further studied in~\cite{Non,Non2}. As we will see in the sequel, from such curved anisotropic  $1:2$  oscillator the curved RDG series  $ \mathcal{V}_{\k,n}$  can be fully constructed in a self-consistent way, and the corresponding curved H\'enon--Heiles system will be obtained as the superposition of the $n=2$ and $n=3$ terms from this series. Therefore, let us recall the explicit form of such curved $1:2$  anisotropic oscillator.

\medskip

\noindent
{\bf Proposition 1.}  \cite{RS,Non} {\em  Let  $\nonn^{1:2}_\kk$ be  the Hamiltonian      
\be
\nonn^{1:2}_\kk= {\cal T}_\k+ \mathcal{V}_\k^{1:2 }=
\frac 12 \left(1+\kk \bq^2\right) \left(\bp^2+\kk(\bq\cdot\bp)^2 \right) +\Om\, \frac{  q_1^2(1+\k q_2^2)  + 4q_2^2}{(1-\kk  \tq_2^2)^2  } ,
 \label{dc}
    \ee
which is written in terms of  Beltrami coordinates~\eqref{Beltr} and their conjugate momenta.
This system is superintegrable for any value of $\k$ and $\Omega$, and the    two functionally independent integrals of motion for $\nonn^{1:2}_\kk$ are 
\be   
   \iii^{1:2}_{\kk}= \frac{1}{2} \left( J_{01}^2+\kk J_{12}^2 \right)   + \Om\, \frac{\tq_1^2(1+\kk \tq_2^2) }{(1-\kk \tq_2^2)^2} ,
\qquad 
 \ele_{\kk}^{1:2}= J_{01} J_{12}  +2\Omega\,   \frac{\tq_1^2\tq_2}{(1-\kk \tq_2^2)^2 } ,
\nonumber
          \ee 
         where  $J_{01}$,  $J_{12}$   are the functions   given by (\ref{cd}). }

\medskip

In the Euclidean space with $\k=0$ the above expressions are reduced to the well-known flat ones, namely
\bea
&& \nonn^{1:2}= {\cal T} +  \mathcal{V}^{1:2} = \frac{1}{2}\, 
\>p^2+\Om \left( q_{1}^2+ 4q_{2}^2 \right) ,\nonumber\\[2pt]
 &&  \iii^{1:2}= \frac{1}{2}\, p_1^2+\Om\,   \tq_1^2  ,\qquad  \ele^{1:2}= p_1  (q_1 p_2 - q_2 p_1)  +2\Omega\,    {\tq_1^2\tq_2} .
\label{ddx}
\eea
Hence $  \iii^{1:2}_{\kk}$ is   the curved generalization of the integral   $ \iii^{1:2}$, which  comes from the separability of  the flat system in Cartesian coordinates,  meanwhile
the additional  integral $ \ele_{\kk}^{1:2}$ ensures superintegrability. 
Note that the expressions (\ref{ddx}) can  be directly  written in terms of the RDG potential (\ref{Ram1}) and its integrals of motion (\ref{Rflat}) with $n=2$ as
 \bea
&& \mathcal{H}^{1:2}\equiv  \mathcal{H}_2 =\frac{1}{2}\, 
\>p^2+\alpha_{2} \mathcal{V}_{2}  ,\qquad      \mathcal{V}^{1:2} \equiv \mathcal{V}_{2}=q_{1}^2+ 4q_{2}^2,
\nonumber\\
&& \mathcal{L}^{1:2}\equiv  \mathcal{L}_2 =  
p_{1}(q_{1}p_{2}-q_{2}p_{1})+\alpha_2  q_{1}^{2} \mathcal{V}_{1} ,
\nonumber
\eea
where $\mathcal{V}_{1} =2q_{2} $ and $\alpha_2=\Om$. 

Our Ansatz will be that this integrability structure in terms of  $\mathcal{V}_{1}$ and $\mathcal{V}_{2}$ should be preserved in the curved case for a suitably defined curved RDG Hamiltonian, $ \mathcal{H}_{\k,n} $. Therefore  such Hamiltonian must be integrable for any value of $n$. With this in mind we can rewrite the expressions from Proposition 1 in the form
 \bea
&&\nonn^{1:2}_\kk\equiv  \mathcal{H}_{\k,2}  = {\cal T}_\k   +\alpha_{2} \mathcal{V}_{\k,2}  ,\qquad \mathcal{V}_{\k,2}=    \frac{  q_1^2(1+\k q_2^2) + 4 q_2^2}{(1-\kk  \tq_2^2)^2  } \, ,
\nonumber\\[2pt]
&& \mathcal{L}^{1:2}_\k\equiv  \mathcal{L}_{\k,2} =  J_{01} J_{12} +\alpha_2 \frac{ q_{1}^{2} } {1+ \kappa  \bf{q}^{2} }  \mathcal{V}_{\k,1} , \qquad 
\nonumber
\eea
so that $\alpha_2 \mathcal{V}_{\k,2}=\mathcal{V}_\k^{ 1:2}$ (\ref{dc}) and where the curved RDG potential with $n=1$ would be
\be
\mathcal{V}_{\k,1}=\frac{2 q_{2}(1+\kappa \textbf{q}^{2})}{(1-\kappa  q_{2}^{2})^{2}} \, .
\nonumber
\ee
Now, by taking into account (\ref{Rflat}) we can also propose the following curved  expression for the RDG Hamiltonian when $n=1$:
  \be
 \mathcal{H}_{\k,1}  = {\cal T}_\k   +\alpha_{1} \mathcal{V}_{\k,1}  , 
 \nonumber
 \ee
 whose integral of the motion is easily shown to be 
 \be
 \mathcal{L}_{\k,1} =  J_{01} J_{12} +\alpha_1 \frac{ q_{1}^{2} } {1+ \kappa  \bf{q}^{2} }  \mathcal{V}_{\k,0}.
 \nonumber
 \ee
where  the curved version of the constant ($0$-th order) RDG potential $ \mathcal{V}_{0}=1$ would be
 \be
  \mathcal{V}_{\k,0}:=\dfrac{(1+ \kappa q_{2}^{2})(1+ \kappa \>q^2  )}{\left(1- \kappa q_{2}^{2}\right)^{2}} \, .
\label{fd}
\ee
Surprisingly enough, this means that the self-consistency conditions that guaranteee the existence of a curved RDG series lead to the fact that $ \mathcal{V}_{\k,0}$ is no longer a constant function. In fact, from these results the form of the full curved RDG series for arbitrary $n$ can be  obtained by induction, and is the following:

 \medskip
\noindent
{\bf Proposition 2.}  {\em  The  curved RDG potentials on   the sphere    ${\>S}^{2}$ and   the  hyperbolic space  ${\>H}^{2}$ are  defined in Beltrami coordinates by
\begin{equation}
\mathcal{V}_{\k, n} =\left(\dfrac{1+\kappa   {\>q}^{2}}{1-\k q_{2}^{2}}\right)^{2}
\sum\limits_{i=0}^{[\frac{n}{2}]}2^{n-2i}\dbinom{n-i}{i}  \! \left(
\dfrac{q_{1}}{\sqrt{1+\kappa   {\>q}^{2}}}
\right)^{2i} \!\! \left(
1-\dfrac{i }{n-i}\left[\dfrac{\kappa  q_{1}^{2}}{1+\kappa   {\>q}^{2}}\right]
\right)
\!\left(
\dfrac{q_{2}}{1+\kappa   {\>q}^{2}}
\right)^{n-2i}
\label{fe}
\end{equation}
with $n\in\mathbb{N}^+$. The curved RDG Hamiltonian
\be
 \mathcal{H}_{\k,n}  = {\cal T}_\k   +\alpha_{n} \mathcal{V}_{\k,n} \, , 
\nonumber
 \ee
is integrable with a constant of motion quadratic in the momenta, since the function
\begin{equation}
\mathcal{L}_{\k,n} =J_{01}J_{12}+\alpha_{n} \dfrac{q_{1}^{2}}{1+ \kappa \bf{q}^{2} } \mathcal{V}_{\k, n-1}\, , 
\nonumber
\end{equation}
fulfils $\{\mathcal{H}_{\kk,n} ,\mathcal{L}_{\kk,n} \}=0$.  
Here $\mathcal{V}_{\k, 0} $ is given by  (\ref{fd}) and  $J_{01}$,  $J_{12}$  and  ${\cal T}_\k$  are the functions   given by (\ref{cd}) and (\ref{ce}).}
\medskip

Notice that under the flat limit $\kappa\rightarrow 0$     the above expressions  straightforwardly lead to the Euclidean   $\mathcal{V}_{n}$ potential (\ref{Ram1}), 
and to $\mathcal{H}_{ n}$ and $\mathcal{L}_{n}$ (\ref{Rflat}) together with  $\mathcal{V}_{0}=1$.  
On the other hand, the curved RDG potentials  (\ref{fe})  can also be straightforwardly written in terms of the  ambient coordinates $(x_0,\>x)$ by applying (\ref{Beltr}). They read
\be
\mathcal{V}_{\k,n} =\dfrac{1}{(x_{0}^{2}-\kappa x_{2}^{2})^{2}}\sum\limits_{i=0}^{[\frac{n}{2}]}2^{n-2i}\dbinom{n-i}{i}x_{1}^{2i}\left(
1-\dfrac{i}{n-i}\kappa  x_{1}^{2}
\right) \left(x_{0} x_{2}\right)^{n-2i} ,\quad n\in\mathbb{N}^+ .
\label{RSerie}
\ee
This expression facilitates the translation of such potentials in any other coordinate system. 
In Table 1 we illustrate all these results for $n=0,1,\dots,4$.  Recall that when $\k=0$ the ambient coordinates are just $x_0=1$ and  $(x_1,x_2)=(q_1,q_2)$.


\begin{table}[t]

 {\footnotesize{

\caption{{The    potentials in the RDG series  for $n=0,1,\dots,4$ on ${\mathbf E}^2$ in Cartesian coordinates $\>q$ (\ref{Ram1}) and  their curved counterpart on  ${\mathbf S}^2$ and  ${\mathbf H}^2$ in   ambient coordinates (\ref{RSerie}) such that  $x_{0}^{2}+\kappa \>x^2=1$. }}
\label{Table1}
 \begin{center}
\noindent
\begin{tabular}{llll}
\hline
\\[-0.2cm]
\multicolumn{1}{l}{ ${\mathbf E}^2$: Cartesian coordinates $\>q$}&
\multicolumn{1}{l}{\qquad${\mathbf S}^2$ and  ${\mathbf H}^2$:  Ambient coordinates $(x_0,\>x)$}
 \\[0.2cm]
\hline
 \\[-0.2cm]
$ \mathcal{V}_{0}=1 $&\qquad$ \displaystyle{ \mathcal{V}_{\k,0} =\dfrac{1- \kappa x_{1}^{2}}{(x_{0}^{2}-\kappa x_{2}^{2})^{2}}}  $\\[10pt]
$ \mathcal{V}_{1}=2 q_2 $&\qquad$ \displaystyle{  
 \mathcal{V}_{\k,1} =\dfrac{2 x_0x_2  }{(x_{0}^{2}-\kappa x_{2}^{2})^{2}}}  $\\[10pt]
$ \mathcal{V}_{2}=q_{1}^2+ 4q_{2}^2   $&\qquad$ \displaystyle{  
 \mathcal{V}_{\k,2}=   \frac{x_1^2 (1-\k x_1^2)+  4 x_0^2x_2^2  }{(x_{0}^{2}-\kappa x_{2}^{2})^{2}}}  $\\[10pt]
$ \mathcal{V}_{3}=4 q_{1}^2 q_2+ 8q_{2}^3   $&\qquad$ \displaystyle{  
 \mathcal{V}_{\k,3}=   \dfrac{ 4 x_0 x_1^2  x_2 (1-\frac 12 \k x_1^2)+ 8  x_0^3x_2^3  }{(x_{0}^{2}-\kappa x_{2}^{2})^{2}}}  $\\[10pt]
$ \mathcal{V}_{4}=  q_{1}^4+ 12 q_{1}^2 q_{2}^2  + 16 q_{2}^4    $&\qquad$ \displaystyle{  
 \mathcal{V}_{\k,4}=   \dfrac{   x_1^4  (1-   \k x_1^2) + 12 x_0^2 x_1^2  x_2^2 (1-\frac 13 \k x_1^2)+ 16  x_0^4x_2^4  }{(x_{0}^{2}-\kappa x_{2}^{2})^{2}}}  $\\[10pt]
$\qquad\qquad   {\vdots}$ &\quad$\qquad \quad {\vdots}$ \\[4pt]
\hline
\end{tabular}
\end{center}
}}
 \end{table}


The next step consists in demonstrating that, as in the Euclidean  case, the curved RDG potentials can be  freely superposed, so generalizing to the curved case the expressions (\ref{bk}) and (\ref{I2Ramani}). This result can be obtained through straightforward computations and reads:
  
 \medskip
\noindent
{\bf Theorem 3.}  {\em For any $M\in\mathbb{N}^+$, the Hamiltonian formed by the  superposition of   the curved RDG potentials (\ref{fe}) and given by
\begin{equation}
\mathcal{H}_{\kappa,\M M}= {\cal T}_\k +\sum\limits_{n=1}^{M}\alpha_{n}\mathcal{V}_{\k,n} \, ,
\nonumber
 \end{equation}
Poisson-commutes with  the function
\begin{equation}
\mathcal{L}_{\kappa,\M M}= J_{01}J_{12}+\dfrac{q_{1}^{2}}{1+ \kappa  \bf{q}^{2}}    \sum\limits_{n=1}^{M}\alpha_{n}\mathcal{V}_{\k, n-1}  .
\nonumber
\end{equation}
}

 \medskip

Consequently,  we have obtained all the ingredients needed in order to obtain the curved counterpart of the    H\'enon--Heiles Hamiltonian (\ref{KdVflat}) and its integral  (\ref{I2KdVflat}), 
since  Proposition 2 provides the new curved `cubic'   potential $\mathcal{V}_{\k,3}$, while  Theorem 3 establishes the appropriate integrable superposition with the known
curved `quadratic' term  $\mathcal{V}_{\k,2}$. In this way  we find,   through the identification of parameters given by (\ref{xz}), the following main result of the paper, which gives the expression for an integrable curved H\'enon--Heiles Hamiltonian of KdV type that preserves the full integrability structure of the flat Hamiltonian system.


 \medskip
\noindent
{\bf Proposition 4.}  {\em The curved counterpart of the H\'enon--Heiles Hamiltonian~\eqref{KdVflat} on ${\mathbf S}^2$ and ${\mathbf H}^2$ is written in Beltrami variables  as
\bea
&& \mathcal{H}_\kk  ={\cal T}_\k + {\cal V}_\k={\cal T}_\k +\alpha_{2} \mathcal{V}_{\kk, 2}+\alpha_{3} \mathcal{V}_{\kk, 3} 
\nonumber\\[2pt]
&&   \quad\  \ = \frac 12 \left(1+\kk \bq^2\right) \left(\bp^2+\kk(\bq\cdot\bp)^2 \right) +\Om\,   \frac{  q_1^2(1+\k q_2^2) + 4 q_2^2}{(1-\kk  \tq_2^2)^2  }\nonumber\\[2pt]
&&\qquad\qquad \quad  +
 {\alpha}\,  \frac{   q_1^2 q_2(1+\k\>q^2- \frac 12 \k q_1^2)  + 2q_2^3}{ (1-\kk  \tq_2^2)^2 (1+\k \>q^2)  } 
\label{bex1}\nonumber
\eea
and Poisson-commutes with the function
\bea
&&  \mathcal{I}_\kk =  J_{01}J_{12}+\frac{q_{1}^{2}}{1+ \kappa  \bf{q}^{2}} \left(\alpha_2\mathcal{V}_{\k,1} +
 {\alpha_3} \mathcal{V}_{\k,2}   
\right)\nonumber\\[2pt]
&&\quad\ =   \left(   p_1+\kk (\bq\cdot\bp) q_1  \right)\left( q_1 p_2 - q_2 p_1\right)\nonumber\\[2pt]
&&\qquad \qquad
+\frac{q_{1}^{2}}{1+ \kappa  \bf{q}^{2}} \left(\Omega \frac{2 q_{2}(1+\kappa  {\>q}^{2})}{(1-\kappa  q_{2}^{2})^{2}}+
 {\alpha} \, \frac{  q_1^2(1+\k q_2^2) + 4 q_2^2}{4(1-\kk  \tq_2^2)^2  }  
\right)\, ,
\label{bex2}\nonumber
\eea
such that $\alpha_2=\Omega$ and   $\alpha_3= \alpha/4$.
}
\medskip

Of course, the vanishing curvature $\kappa\to 0$ limit of these two expressions is, respectively,~\eqref{KdVflat} and~\eqref{I2KdVflat}. Moreover, Theorem 3 can now be  understood as the tool that provides further integrable deformations of this curved H\'enon--Heiles system whose flat limit is a polynomial potential of degree $M$. On the other hand, the expression of the curved H\'enon--Heiles potential ${\cal V}_\k$ in terms of ambient coordinates can be easily derived from Table 1.


\sect{Concluding remarks and open problems}

In this paper we have presented the generalization on 2D spaces with constant curvature of the $1:2$ integrable H\'enon--Heiles system of KdV type. This new Liouville integrable Hamiltonian system has been constructed by following a `curved integrability criterium', {\em i.e.}, by exploring all the integrability properties of the Euclidean system $ \mathcal{H}$ and by constructing its curved analogue $ \mathcal{H}_\kappa$ by preserving the full integrability structure of the flat case. In this way, by starting from the known $1:2$ curved superintegrable anharmonic oscillator, the full curved RDG series of integrable potentials have been constructed, and the H\'enon--Heiles system can thus be  obtained as the appropriate superposition of the second- and third-order curved RDG potentials. We stress that the fact that the latter series of potentials can be expressed as rational functions of the (projective) Beltrami coordinates turns out to be very helpful from the computational viewpoint (see also~\cite{Albouy} and references therein for a recent review on the various facets of the projective dynamics approach).

This result suggests in a natural way several open problems that we plan to face in the near future through the same approach. The first one would be the construction of the curved analogue of the H\'enon--Heiles system of KdV type with arbitrary anisotropic oscillator frequencies ($\beta=2$ and $(\Omega_{1},\Omega_{2})$ arbitrary in~\eqref{hhmulti}), as well as the study of the generalization to the curved case of the integrable rational perturbations of the KdV flat H\'enon--Heiles system introduced in~\cite{FF,HonePLA,tesis,Annals10}. 

The following natural step would  consist in the construction of  the curved Sawada--Kotera H\'enon--Heiles system  (\ref{wqa}) ($\beta=1/3$ and $\Omega_{1}=\Omega_{2}$) by starting from the also known superintegrable  curved $1:1$ isotropic (Higgs) oscillator (see~\cite{Non, Higgs}). Since the flat Sawada--Kotera system is separable in rotated Cartesian coordinates, an associated series of integrable polynomial perturbations does also exist, and the curved Sawada--Kotera system should be obtained by following a similar procedure to the one presented in this paper. In the same manner,  the curved Kaup--Kuperschdmit H\'enon--Heiles system  (\ref{wqc})   ($\beta=16/3$ and $\Omega_{2}=16\,\Omega_{1}$) should be obtained by starting from a  superintegrable  curved $1:4$ oscillator whose integral of the motion should be quartic in the momenta, but in this case even the latter curved anharmonic oscillator is still unknown. Finally, the construction of the quantum integrable version of all these new curved systems as well as the study of their quantum dynamics constitute a challenging problem. In this respect, we recall that the preservation of the integrability properties under the deformation induced by the curvature turns out to be essential in order to guarantee the exact solvability of the corresponding quantum system~\cite{annalssuper}.


\section*{Acknowledgements}

This work was partially supported by the Spanish Ministerio de Econom\1a y Competitividad     (MINECO) under grants MTM2010-18556 and MTM2013-43820-P, and by the Spanish Junta de Castilla y Le\'on  under grant BU278U14.   The authors acknowledge the referees for their valuable reports.


\end{document}